\documentclass[11pt,preprintnumbers,aps,amssymb,nofootinbib,amsmath,superscriptaddress]
{revtex4}
\usepackage{epsfig,epsf}
\usepackage{bm} % puts greek and math symbols in boldface using \bm
\usepackage{slashed}
\usepackage{color} % {\color{red} ... }
\newcommand{\beq}{\begin{equation}}
\newcommand{\beql}[1]{\begin{equation}\label{#1}}
\newcommand{\eeq}{\end{equation}}
\def\bal#1\gal{\begin{align}#1\end{align}}
\newcommand{\ball}[1]{\bal\label{#1}}
\newcommand{\eq}[1]{(\ref{#1})}
\newcommand{\fig}[1]{Fig.~\ref{#1}}
\renewcommand{\sec}[1]{Sec.~\ref{#1}}
\newcounter{topiccounter}
\renewcommand{\b}[1]{{\bm #1}} 
\newcommand{\aver}[1]{\left\langle #1 \right\rangle}
\begin{document}

\title{Time-evolution of magnetic field in hot nuclear matter with fluctuating topological charge }
\author{Kirill Tuchin}

\affiliation{Department of Physics and Astronomy, Iowa State University, Ames, Iowa, 50011, USA}

\date{\today}


\begin{abstract}
 
The time-evolution of the magnetic field in hot homogeneous nuclear matter has two qualitatively different stages separated by the sphaleron transition time $\tau_c$. At early times, when the chiral conductivity $\sigma_\chi$ is a slow function of time, the soft chiral modes $k<\sigma_\chi$ of the magnetic field  grow exponentially  with time, which is known as the chiral instability. At later times $\sigma_\chi$ fluctuates due to the  sphaleron transitions and can be regarded as a random process. It is argued that the average magnetic field is exponentially damped at later times. The time-evolution of the average field energy is more complicated and depends on the electrical conductivity of the chiral matter but does not depend on chirality. It exhibits instability only if the matter is a poor electrical conductor, such as the quark-gluon plasma near  the critical temperature. The precise conditions for the instability and the growth rate of the unstable modes are derived.    

\end{abstract}

\maketitle

%%%%%%%%%%%%%%%%%%%%%%%%%%%%%%%%%%%%%%%%
\section{Introduction}\label{sec:i}

The topological configurations of gluon fields  form $P$ and $T$-odd domains in hot nuclear matter. Their topological charge couples to the electromagnetic fields by way of the chiral anomaly.  Electrodynamics, modified in the presence of the topological domains, exhibits a number of novel effects such as the chiral magnetic effect \cite{Kharzeev:2007jp,Fukushima:2008xe} and the anomalous Hall effect \cite{Haldane:2004zz}. One of its features that attracted a considerable interest is the emergence of the soft magnetic field modes exponentially growing in time known as the chiral instability \cite{Tuchin:2014iua,Manuel:2015zpa,Joyce:1997uy,Boyarsky:2011uy,Kharzeev:2013ffa,Khaidukov:2013sja,Kirilin:2013fqa,Avdoshkin:2014gpa,Akamatsu:2013pjd,Dvornikov:2014uza,Buividovich:2015jfa,Sigl:2015xva,Xia:2016any,Kaplan:2016drz,Kirilin:2017tdh,Mace:2019cqo}. These unstable modes transfer helicity from the chiral medium, such as the quark-gluon plasma, to the magnetic field in a process known as the inverse cascade \cite{Biskamp,Boyarsky:2011uy}.

The chiral instability can be most easily derived using the effective Maxwell-Chern-Simons theory  which adds to the Maxwell Lagrangian a term that couples $F\tilde F$ directly to the topological domains by means of the chiral anomaly \cite{Wilczek:1987mv,Carroll:1989vb,Sikivie:1984yz}. The dynamics of the electromagnetic field in the spatially uniform chiral matter with electrical conductivity $\sigma$ can be described by the vector potential $\b A$ which satisfies the following equation in the radiation gauge 
\ball{m1}
-\nabla^2 \b A= -\partial_t^2\b A-\sigma \partial_t\b A  +\sigma_\chi(t) \b \nabla\times \b A\,,
\gal
where $\sigma_\chi$ is the chiral conductivity \cite{Kharzeev:2009pj,Fukushima:2008xe} sourced by the topological gluon field configurations, see Appendix~A for more details. The ratio $\mu_5= \sigma_\chi/c_A$, where $c_A$ is the anomaly coefficient,  is also used throughout the paper.\footnote{In models with fermions $\mu_5$ can be identified as the axial chemical potential. However, 
its physical interpretation is not free of difficulties, see \cite{Huang:2015oca} and references therein.} Seeking a solution in the form $\b A = a_{\b k \lambda}(t)\b \epsilon_\lambda e^{i\b k \cdot \b x}$, where $\b \epsilon_\lambda$ are the circular polarization vectors of helicity $\lambda=\pm 1$ and using the identity $i\b k  \times \b \epsilon_\lambda = k\lambda \b \epsilon_\lambda$ one deduces that  the amplitude $a_{\b k\lambda}$ satisfies the equation
\ball{m3}
\ddot a_{\b k \lambda} + \sigma \dot a_{\b k \lambda} + k(k- \lambda\sigma_\chi) a_{\b k \lambda}=0\,.
\gal
Assuming that the chiral conductivity is a slow function of time  and neglecting for the sake of brevity the electrical conductivity, the magnetic field for each plane wave mode reads, up to the pre-exponential factors, 
\ball{k2}
\b B  \sim e^{i\b k\cdot \b x }\exp\left\{\mp it\sqrt{k(k+\lambda\sigma_\chi)}\right\}\,.
\gal
 Clearly, when $\lambda \sigma_\chi<0$ the magnetic field modes  $k<|\lambda\sigma_\chi|$ are unstable \cite{Tuchin:2014iua}. This instability is manifestly chiral. 
 
It is remarkable, that  since \eq{k2} satisfies the modified Maxwell equations, the exponential growth of the magnetic field is not constrained by the energy conservation \cite{Kaplan:2016drz,Tuchin:2017vwb}. However, it is constrained by the conservation of the total helicity of the magnetic field and the chiral matter, which follows from the chiral anomaly equation \cite{Adler:1969gk,Bell:1969ts}. As the magnetic field increases, the helicity flows from the chiral matter to the field until the chiral conductivity vanishes \cite{Boyarsky:2011uy,Hirono:2015rla,Kaplan:2016drz,Tuchin:2017vwb}. As argued in \cite{Tuchin:2017vwb},
this helicity transfer is an adiabatic process, meaning that $\sigma_\chi$ is a slowly varying function of time. 

Let us now examine the assumptions that led us from a MCS equation \eq{m1} to the unstable solution \eq{k2}. Dropping the first assumption $\sigma=0$ results in a more complicated expression for the magnetic field. However, the finite value of the electrical conductivity does not have any significant effect upon the instability dynamics at constant (or slowly varying) $\sigma_\chi$ \cite{Tuchin:2017vwb}. The second and crucial assumption is the slow variation of the chiral conductivity. 
It holds while $t<\tau_c$ where $\tau_c$ is the sphaleron transition time, i.e.\ the transition time between the gauge field configurations of different topological charge. By the time $t=\tau_c$, the  exponent in \eq{k2} reaches its maximal value $\exp(\sigma_\chi \tau_c)$. The estimates  of the sphaleron transition time $\tau_c\sim 1/(g^4T)$ \cite{Arnold:1996dy} and the chiral conductivity $\sigma_\chi \sim e^2\mu_5$ \cite{Fukushima:2008xe,Kharzeev:2009pj}, at temperature $T$,  indicate that $\sigma_\chi \tau_c\ll 1$\footnote{This holds true since $e^2/g^4\ll 1 $ at any reasonable temperature and $\mu_5$ is expected to be of the order of $T$ or smaller. A more accurate estimate for the sphaleron transition time is $\tau_c\sim 1/(g^4T)\ln g^{-1}$  \cite{Bodeker:1998hm,Arnold:1998cy}.}. The  implication is that during the time interval $t<\tau_c$ when the chiral conductivity (and the corresponding topological charge density) may be regarded as slowly varying functions of time, the magnetic field instability simply does not have enough time to grow.
Thus, in order to study the magnetic field instability in the hot nuclear matter, one is required to examine the opposite limit of $t\gg \tau_c$, when many sphaleron transition can occur causing fluctuations of the topological charge of the chiral domains.

The main goal of this paper is to examine this limit by regarding the chiral conductivity as a  stochastic process with vanishing expectation value $\aver{\sigma_\chi}=0$. I assume that its dispersion $\Sigma_\chi=\sqrt{\aver{\sigma_\chi^2}}$ equals $c_A\mu_5$ and the auto-correlation function $\aver{\sigma_\chi(t)\sigma_\chi(t-\tau)}$ has support at $t<\tau_c$, meaning that the sphaleron transition time $\tau_c$ is the correlation time. Eq.~\eq{m3} then describes the harmonic oscillator with a random frequency. Using the Van Kampen's theory  \cite{VanKampen:1975} one can deduce from \eq{m3} the ordinary differential equations for the average values of the field amplitude $\aver{a_{\b k\lambda}}$ and its second moments $\aver{(a_{\b k\lambda})^2}$, $\aver{(\dot a_{\b k\lambda})^2}$. The main result of this theory is 
that the average value of the magnetic field decays exponentially with time. The average value of the magnetic field energy also decays if the plasma is a good electric conductor, which occurs at high temperatures. However, at temperatures close to the critical temperature the plasma is a poor conductor and the average energy is unstable.

An important feature of the chiral instability is its remarkable universality in many chiral systems. The universality emerges in the adiabatic limit, when the details of the topological charge distribution are irrelevant, so that, for example,  electromagnetic plasma with chirality imbalance and the quark-gluon plasma with a topological gluon field configuration have similar chiral instability, described by one parameter -- the chiral conductivity, see \cite{Tuchin:2014iua,Manuel:2015zpa,Joyce:1997uy,Boyarsky:2011uy,Kharzeev:2013ffa,Khaidukov:2013sja,Kirilin:2013fqa,Avdoshkin:2014gpa,Akamatsu:2013pjd,Dvornikov:2014uza,Buividovich:2015jfa,Sigl:2015xva,Xia:2016any,Kaplan:2016drz,Kirilin:2017tdh,Mace:2019cqo} and references therein. This paper extends analysis of the chiral instability to systems whose  topological charge fluctuations are stochastic in the long time limit. This approach is adequate for hot nuclear matter since the topological charge is determined mostly by the strong interactions, but may not universally hold in other chiral systems.

The paper is structured as follows. In \sec{sec:m} the Van Kampen theory is reviewed and employed to reduce the stochastic equation \eq{m3} to the ordinary differential equations for the averages of the amplitude and its second moment. The solutions to these equations are used in \sec{sec:M} and \sec{sec:n} to analyze the time evolution of the average magnetic field, electromagnetic energy and magnetic helicity with time. The summary and discussion is presented in \sec{sec:s}.

%%%%%%%%%%%%%%%%%%%%%%
\section{Evolution equations for the amplitude moments  }\label{sec:m}

To study the late time behavior of the magnetic field, one can regard the chiral conductivity $\sigma_\chi(t)$ as a random process and hence \eq{m3} becomes a stochastic equation describing time-evolution of the field amplitude with momentum $\b k$ and polarization $\lambda$. Eq.~\eq{m3} does not have an analytical solution. However, one can deduce from it ordinary differential equations for the expectation value of the amplitude  moments using the Van Kampen theory  \cite{VanKampen:1975}. This section represents a brief summary of the relevant results. For the reasons explained below, the underdamped   and the overdamped  modes with $k\gtrsim \sigma/2$ and $k\lesssim \sigma/2$ respectively are considered separately from the critically damped ones with $k\approx \sigma/2$.

%%%%%
\subsection{Underdamped and overdamped modes}

To begin with, I define, for a given $\b k$ and $\lambda$,  a new variable $x= a_{\b k \lambda} e^{\sigma t/2}$ which satisfies the equation
\ball{m5}
\ddot x(t) + \omega^2[ 1+ \alpha\xi(t)]x(t)=0\,,
\gal
where 
\ball{m6}
\omega^2= k^2-\frac{\sigma^2}{4}\,,\qquad
\alpha=-\frac{\lambda k}{\omega^2}\Sigma_\chi \,,\qquad
\xi= \frac{\sigma_\chi}{\Sigma_\chi}\,, \qquad \Sigma_\chi= \sqrt{\aver{\sigma^2_\chi}}\,.
\gal
The random process $\xi(t)$ is such that $\aver{\xi}=0$. The correlation time of $\xi$ is $\tau_c$, meaning that the auto-correlation function $\aver{\xi(t)\xi(t-\tau)}$ vanishes at $\tau>\tau_c$. For the modes with $\omega\neq 0$ it is convenient to introduce the dimensionless time variable $t'=\omega t$. 

Eq.~\eq{m5} can be cast in the matrix form
\ball{o1}
\frac{du(t')}{dt'} = [A_0+\alpha \xi(t') B]u(t')\,,
\gal 
where $u=(x, x')^T$ with $x'= dx/dt'$ and  
\ball{m8}
A_0= \left(\begin{array}{cc}0 & 1 \\-1 & 0\end{array}\right)\,,\qquad
B= \left(\begin{array}{cc}0 & 0 \\-1 & 0\end{array}\right)\,.
\gal
 The main result of \cite{VanKampen:1975} is that given Eq.~\eq{o1} with arbitrary $u$, $A_0$ and  $B$, at late times $t\gg \tau_c$ the expectation value of $u$ satisfies the equation
\ball{o3} 
\frac{d\aver{u(t')}}{dt'}= \left\{ A_0+\alpha^2\int_0^\infty \aver{\xi(t')\xi(t'-\tau')} Be^{A_0\tau'}Be^{-A_0\tau'}d\tau'\right\} \aver{u(t')}\,,
\gal
provided that 
\ball{o4}
|\alpha\omega| \tau_c\ll 1\,,
\gal
 which allows treating the fluctuating term in \eq{m5}  as a perturbation.  In view of $\omega$-dependence of $\alpha$ in \eq{m6}, the condition \eq{o4} excludes modes with small $\omega$'s. Namely, expanding $k= \sigma/2+\delta k$ one obtains that \eq{o4} is satisfied if 
 \ball{o5} 
|\delta k|\gg \frac{\sigma}{4}(\Sigma_\chi \tau_c)^2\,.
\gal
Since $\Sigma_\chi\tau_c\sim e^2\mu_5/g^4T\ll 1$, \eq{o4} holds for all $k$'s except in the narrow interval around $k=\sigma/2$. The modes with $\delta k \lesssim \sigma(\Sigma_\chi \tau_c/2)^2$ will be referred to as the  \emph{critically dumped} modes, while those satisfying \eq{o5} with $\delta k>0$ and $\delta k<0$ as \emph{underdamped} and \emph{overdamped} modes respectively.

 Eq.~\eq{o3} can now be applied to the underdamped and overdamped modes. Computing with the help of \eq{m8} 
\ball{m9}
Be^{A_0\tau'}Be^{-A_0\tau'}= \left(\begin{array}{cc}0 & 0 \\ \sin\tau'\cos\tau' & -\sin^2\tau' 
\end{array}\right)\,,
\gal
substituting into \eq{o3} and converting the result into the second order differential equation for $\aver{x}$ yields
\ball{m13}
\frac{d^2\aver{x}}{dt'^2}+\frac{1}{2} \alpha^2 c_2 \frac{d\aver{x}}{dt'}+\left( 1-\frac{1}{2}\alpha^2 c_1\right)\aver{x}=0\,,
\gal
where
\begin{subequations}\label{m15+}
\bal
&c_1=\int_0^\infty  \aver{\xi(t')\xi(t'-\tau')}\sin(2\tau')d\tau'\,,\label{m15}\\
&c_2=\int_0^\infty \aver{\xi(t')\xi(t'-\tau')}[1-\cos(2\tau')]d\tau'\,.\label{m16}
\gal
\end{subequations}

The evolution equations for the second moments of $x$ can be obtained by applying \eq{o1},\eq{o3} to the vector $u=(x^2, x'^2,x x')^T$. The corresponding matrices are
\ball{m18}
A_0= \left(\begin{array}{ccc}0 & 0 & 2 \\0 & 0 & -2 \\ -1 & 1 & 0\end{array}\right)\,,\qquad 
B= \left(\begin{array}{ccc}0 & 0 & 0 \\0 & 0 & -2 \\ -1 & 0 & 0\end{array}\right)
\gal
which produces
\ball{m18+}
Be^{A_0\tau'}Be^{-A_0\tau'}= \left(\begin{array}{ccc}0 & 0 & 0 \\ 2\cos^2\tau' & -2\sin^2\tau' & 0 \\ \sin2\tau' & 0 & -2\sin^2\tau'\end{array}\right)\,.
\gal
Eq.~\eq{o3} now reads
\ball{n1}
\frac{d}{dt'}
\left(\begin{array}{c} \aver{x^2} \\
\aver{x'^2} \\ \aver{x x'}\end{array}\right)
=\left(\begin{array}{ccc}0 & 0 & 2 \\ \alpha^2 c_3 & -\alpha^2 c_2 & -2 \\ -1+\alpha^2c_1 & 1& -\alpha^2c_2\end{array}\right)
\left(\begin{array}{c} \aver{x^2} \\
\aver{ x'^2} \\ \aver{x x'}\end{array}\right)
\gal
where 
\ball{n3}
c_3= \int_0^\infty  \aver{\xi(t')\xi(t'-\tau')}(1+ \cos(2 \tau'))d\tau'\,.
\gal

%%%%%%
\subsection{Critically damped modes}\label{subsec:m}

In general, the Van Kampen theory cannot be applied to all critically damped modes because $\omega^2$ is not negligible as compared to the fluctuating term $\lambda k \Sigma_\chi \xi$ in \eq{m5}. The exception is the mode with momentum $k=\sigma/2$ in which case case $\omega=0$ and \eq{m5} takes form 
\ball{m101}
\ddot x(t)+\beta^2 \xi(t) x(t)=0\,,
\gal
where $\xi$ is defined as in \eq{m6} and 
\ball{m102}
\beta^2= -\Sigma_\chi \lambda k=-\frac{1}{2}\Sigma_\chi \lambda\sigma \,.
\gal
Passing to the dimensionless time variable $t'=\beta t$, one can write \eq{m101} in the form \eq{o1} with $\alpha=1$ and
\ball{m104}
A_0= \left(\begin{array}{cc}0 & 1 \\0 & 0\end{array}\right)\,,\qquad
B= \left(\begin{array}{cc}0 & 0 \\-1 & 0\end{array}\right)\,,
\gal
which implies 
\ball{m105}
Be^{A_0\tau'}Be^{-A_0\tau'}= \left(\begin{array}{cc}0 & 0 \\ \tau' & -\tau'^2 
\end{array}\right)\,.
\gal 
The fluctuating term is a small perturbation of the initial condition if $|\beta| \tau_c\ll 1$. In this case, at $t\gg \tau_c$ the expectation value of $u=(x, x')^T$ obeys \eq{o3} with $\alpha=1$. Thus, \eq{m13} with $\alpha=1$ and $t'=\beta t$ can be used to describe the time evolution of $\omega=0$ mode. The corresponding $c$-coefficients read 
\begin{subequations}\label{m106+}
\bal
&c_1=2\int_0^\infty  \aver{\xi(t')\xi(t'-\tau')}\tau' d\tau'\,,\label{m106}\\
&c_2=2\int_0^\infty \aver{\xi(t')\xi(t'-\tau')}\tau'^2 d\tau'\,.\label{m107}
\gal
\end{subequations}

Similarly, $u=(x,  x'^2,x x')^T$ obeys \eq{o1} with 
\ball{m108}
A_0= \left(\begin{array}{ccc}0 & 0 & 2 \\0 & 0 & 0 \\ 0 & 1 & 0\end{array}\right)\,,\qquad 
B= \left(\begin{array}{ccc}0 & 0 & 0 \\0 & 0 & -2 \\ -1 & 0 & 0\end{array}\right)
\gal
and 
\ball{m109}
Be^{A_0\tau'}Be^{-A_0\tau'}= 2\left(\begin{array}{ccc}0 & 0 & 0 \\ 1 & -\tau'^2 & 0 \\ \tau' & 0 & -2\tau'^2 \end{array}\right)\,.
\gal
The corresponding equations for the second moments has the same form as \eq{n1} with $\alpha=1$, and $c$=coefficients given by  \eq{m106},\eq{m107} and 
\ball{n103}
c_3= 2\int_0^\infty  \aver{\xi(t')\xi(t'-\tau')}(1- \tau'^2)d\tau'\,.
\gal

%%%%%%%%%%%%
\subsection{Coefficients $c_1$, $c_2$, $c_3$}

It is useful for future reference  to estimate the coefficients $c_1$, $c_2$, $c_3$ appearing in the evolution equations in the previous subsections. Their $\omega$-dependence is qualitatively different for  good and poor electric conductors.  The hot nuclear matter is a good electric conductor at high temperatures. Its electrical conductivity is 
of the order $\sigma\sim T/e^2$\footnote{More precisely $\sigma=13\, T/e^2\ln e^{-1}$ \cite{Arnold:2000dr}.}.
However, at temperatures above but close to the critical temperature the hot nuclear matter is a poor conductor with the electrical conductivity $\sigma \sim e^2T$ \cite{Aarts:2007wj,Ding:2010ga,Amato:2013oja,Cassing:2013iz,Yin:2013kya}. 

%%%
\subsubsection{Well conducting chiral matter}

(a) Consider the underdamped and overdamped modes satisfying \eq{o5}, which excludes modes in a vicinity of  $k=\sigma/2$. In both cases $|\omega|\tau_c > \sqrt{\sigma \delta k} \tau_c\gg \sigma \Sigma_\chi \tau_c^2 \sim \mu_5/g^8 T\gg 1$ for the electrical conductivity  $\sigma\sim T/e^2$. Since the autocorrelation function equals one at $\tau\ll \tau_c$ and vanishes at $\tau >\tau_c$ one can estimate \eq{m15+},\eq{n3} as
\begin{subequations}\label{p1+}
\bal
&c_1=\omega\int_0^\infty  \aver{\xi(t)\xi(t-\tau)}\sin(2\omega\tau)d\tau\sim \omega\int_0^{1/2\omega}
2\omega \tau d\tau = \frac{1}{4}\,, \label{p1}\\
&c_2 = 2\omega\int_0^\infty  \aver{\xi(t)\xi(t-\tau)}\sin^2(\omega \tau)d\tau 
\sim \omega \int_{0}^{\tau_c} d\tau = \omega\tau_c\,, \label{p2}\\
& c_3 = 2\omega\int_0^\infty  \aver{\xi(t)\xi(t-\tau)}\cos^2(\omega \tau)d\tau = c_2+\mathcal{O}\left(1/ \omega\tau_c\right)\,. \label{p3}
\gal
\end{subequations}
This implies that  $c_2\approx c_3\gg c_1$. 

The coefficient $c_2$ may become negative if the spectrum of $\xi$ contains a mode oscillating with frequency $2\omega$. Indeed the contribution of such mode to $c_2$ can be estimated as 
\ball{p4}
\Delta c_2 \propto \omega\int_0^\infty  \aver{\xi(t)\xi(t-\tau)}[1-\cos(2\omega \tau)]\cos(2\omega \tau)d\tau 
\sim -\frac{\omega}{2} \int_{0}^{\tau_c} d\tau = -\omega\tau_c\,.
\gal
This causes the parametric resonance, i.e.\ the exponential growth of $\aver{x}$ with time, see \eq{m19} below. 

As a specific example, consider the Ornstein-Uhlenbeck random process with the auto-correlation function 
\ball{p6}
\aver{\xi(t)\xi(t-\tau)}=e^{-\tau/\tau_c}\,.
\gal
The corresponding coefficients read
\ball{p7}
c_1&= \frac{2(\omega \tau_c)^2}{1+4(\omega\tau_c)^2}\,,\qquad
c_2 = \frac{4(\omega \tau_c)^3}{1+4(\omega\tau_c)^2}\,,\qquad
c_3= \frac{\left[2+4(\omega \tau_c)^2\right](\omega \tau_c)}{1+4(\omega\tau_c)^2}
\gal
Multiplying \eq{p6} by $\cos(2\omega\tau)$ one can check that $c_1\to 1/8$, $c_2\to -\omega\tau_c/2$ and $c_3\to \omega\tau_c/2$ at $\omega\tau_c\gg 1$. In the next subsection it is shown that at small $|\omega|\tau_c$, the coefficient $c_2$ is always positive and so there is no parametric resonance. 

(b) The Van Kampen theory fails to describe the critically damped modes of the well-conducting plasma because $|\beta|\tau_c\sim \sqrt{\mu_5/T}/g^4\gg 1$ (unless $\mu_5$ is very small). 

%%%
\subsubsection{Poorly conducting chiral matter}

(a) The parameter $\omega\tau_c$ of the underdamped modes $k\gg \sigma/2$ of a poorly conducting plasma can be estimated as $\omega\tau_c\sim k/g^4T$. The coefficients of the modes with $k\gg g^4T$ are given by \eq{p1+}, while the ones with $k\ll g^4T$ are estimated from \eq{m15+} and \eq{n3} as 
\ball{p9}
c_1\sim (\omega\tau_c)^2\,,\qquad c_2\sim (\omega\tau_c)^{3}\,,\qquad c_3\sim \omega\tau_c\,.
\gal
Estimates \eq{p9} apply also to the overdamped modes $k\ll \sigma/2$ because $|\omega|\tau_c<\sigma \tau_c/2\sim e^2/g^4\ll 1$. Incidentally,  this implies that the coefficients \eq{p9}  are strongly ordered as $c_2\ll c_1 \ll c_3\ll 1$.

(b) The $c$-coefficients for the critically damped mode  $k= \sigma/2$ can be  estimated using \eq{m15+},  \eq{n3} with the result similar to \eq{p9}
\ball{n105}
c_1\sim (\beta\tau_c)^2\,,\qquad c_2\sim (\beta\tau_c)^{3}\,,\qquad c_3\sim \beta\tau_c\,.
\gal
In contrast to the well conducting matter,  the Van Kampen theory is valid in this case since $|\beta|\tau_c\sim (e^2/g^4) \sqrt{\mu_5/T} \ll 1$.

%%%%%%%%%%%%%%%%%%%%%%%%%%%%%
%\section{Time-evolution of average magnetic field, its energy and helicity}

%\subsection{Stability of average magnetic field}

\section{Time-evolution of average magnetic field}\label{sec:M}

In this and the following sections the results of the previous section are employed to analyze the late-time behavior of the average magnetic field and average field energy. 

The general solution of \eq{m13} is a linear combination of the functions  
\ball{m19}
\aver{x(t)}_\pm= \exp\left\{\pm i \omega t -\frac{\alpha^2}{4}(c_2\pm i c_1)\omega t\right\} \,,
\gal
where only terms of the order $\alpha^2$ are retained.  The corresponding magnetic field amplitudes are 
\ball{m20}
\aver{a_{\b k \lambda}(t)}_\pm= \exp\left\{\pm i \omega t -\frac{\alpha^2}{4}(c_2\pm i c_1)\omega t-\frac{1}{2}\sigma t\right\} \,.
\gal

%%%%%%
\subsection{The underdamped modes $k> \sigma/2$}

The condition for the instability of the underdamped modes 
\ball{m21.1}
-\frac{\alpha^2}{4}c_2\omega-\frac{\sigma}{2}>0
\gal
can  be satisfied only for the parametric resonance modes with $c_2<0$. Using the definition of $\alpha$ from \eq{m6} this yields  
\ball{m21.2}
k^2\Sigma_\chi^2 |c_2|> 2\omega^3\sigma
\gal 
As mentioned below \eq{p7}, the parametric resonance occurs in good conductors at $\omega\tau_c\gtrsim 1$. Using the estimate  \eq{p4} in \eq{m21.2}  and noting that  the best chance to satisfy it is at smaller $\omega$, expand it near $k=\sigma/2$ to obtain $\delta k< \Sigma_\chi^2\tau_c$. On the other hand, $\omega \gtrsim 1/\tau_c$ implies that $\delta k \gtrsim 1/\sigma\tau_c^2$. The two conditions contradict each other since $\Sigma_\chi^2\sigma \tau_c^3<1$ . Thus, there are no unstable underdamped modes of the average magnetic field. Even the parametric resonance mode in a well-conducting chiral matter is rapidly diffused due to the large electric conductivity.

%%%%%%
\subsection{The overdamped modes $k< \sigma/2$}

The amplitude of the overdamped modes can be written down  as a linear combination of the functions
\ball{m22} 
\aver{a_{\b k \lambda}(t)}_\pm= \exp\left\{\mp  |\omega| t-\frac{ik^2\Sigma_\chi^2}{4\omega^4}(c_2\pm ic_1)|\omega| t-\frac{1}{2}\sigma t\right\}\,.
\gal
The second term in the exponent is a small correction to the first one. Clearly, $\aver{a_{\b k \lambda}(t)}_+$ is stable. In $\aver{a_{\b k \lambda}(t)}_-$ the first and the third terms in the exponent cancel out at $k\ll \sigma/2$. Expanding to the second order in $k/\sigma$ yields 
\ball{m24}
\aver{a_{\b k \lambda}(t)}_-= \exp\left\{-\frac{k^2 t}{\sigma} - \frac{k^2\Sigma_\chi^2 \sigma t(ic_2+c_1)}{8\omega^4}\right\}\,.
\gal
Thus, the overdamped modes are stable at any conductivity. 

%%%%%%
\subsection{The critically damped mode $k= \sigma/2$}

It remains to examine the critically damped mode $\omega= 0$. As noted below \eq{p7} and \eq{n105}  only poor conductors can be consistently considered in the present framework. As shown in \sec{subsec:m}, the average amplitude can be obtained from \eq{m20} by replacing $\omega\to \beta$ and $\alpha\to 1$:
\ball{m26} 
\aver{a_{\b k \lambda}(t)}_\pm= \exp\left\{\pm i \beta t -\frac{1}{4}(c_2\pm i c_1)\beta t-\frac{1}{2} \sigma t\right\}\,.
\gal
The amplitude \eq{m26} of the critically damped mode  exponentially decays at late time because $c_2>0$ as indicated by \eq{n105}.

 %%%%%
 \subsection{Conclusion}

It follows from the analysis in this section that all magnetic field modes are stable at long times $t\gg \tau_c$. 
This of course also includes the modes with $k<\sigma_\chi/2$ that exhibit exponential growth at early times.

 %%%%%%%%%%%%%%%%%%
%%%%%%%%%%%%%%%%%%%
\section{Time-evolution of magnetic field energy and helicity}\label{sec:n}
 
Having examined the time-evolution of the magnetic field which is determined by the first moment of the amplitudes, we turn to the evolution of the electromagnetic energy and magnetic helicity which is determined by the second moments. The electromagnetic field energy and magnetic helicity read 
\bal
&\mathcal{E}= \frac{1}{2}\int (\b E^2+\b B^2)d^3x = \sum_{\b k,\lambda} \mathcal{E}_{\b k\lambda}= \sum_{\b k,\lambda}\frac{1}{2k}\left(\dot a_{\b k \lambda}^2+a_{\b k \lambda}^2k^2\right)\,,\label{n-4}\\
&\mathcal{H}= \int \b A\cdot \b B d^3x =  \sum_{\b k,\lambda}\mathcal{H}_{\b k \lambda}= \sum_{\b k, \lambda}\lambda a_{\b k \lambda}^2\,, \label{n-3}
\gal
%while the energy loss due to Ohm’s currents is
%\ball{n-2}
%Q=\sigma \int \b E^2 d^3x = \sigma\sum_{\b k,\lambda}\frac{1}{k}\dot a_{\b k \lambda}^2\,,
%\gal 
where each mode with given $\b k$ and $\lambda$ is normalized to one particle in a unit volume and the amplitudes $a_{\b k\lambda}$ are real. The corresponding expressions for the expectation values of energy and helicity of each mode can be express in terms of the moments of the variable $x$ as
\bal
&\aver{\mathcal{E}_{\b k\lambda}}= \frac{1}{2k}e^{-\sigma t}
\left[ \aver{x^2}\left(k^2+\frac{1}{4}\sigma^2\right)-\aver{\dot x x}\sigma+\aver{\dot x^2}\right]\,,\label{n13}\\
& \aver{\mathcal{H}_{\b k\lambda}}= \lambda e^{-\sigma t}\aver{x^2}\,. \label{n14}
\gal

The time evolution of the second moments of $x$ is given by \eq{n1}. The eigenvalues of the matrix in \eq{n1} to the order $\alpha^2$ are 
\begin{subequations}\label{n4+}
\bal
&\nu_0= \frac{1}{2}\alpha^2(c_3-c_2)\,,\label{n4}\\
& \nu_\pm = \pm 2i\left(1-\alpha^2\frac{c_1}{4}\right)-\frac{1}{4}\alpha^2 (c_3+3c_2)\,. \label{n5}
\gal
\end{subequations}
Transforming \eq{n1} to the diagonal form and integrating one finds 
\begin{subequations}\label{n7+}
\bal
&\omega^2\aver{x^2}+\aver{\dot x^2}= u_0 e^{\nu_0 \omega t }\,,\label{n7}\\
&\omega^2\aver{x^2}-\aver{\dot x^2}+2i\omega \aver{x\dot x}= 2u_- e^{\nu_- \omega t }\,,\label{n8}\\
-&\omega^2\aver{x^2}+\aver{\dot x^2}+2i\omega \aver{x\dot x}= 2u_+ e^{\nu_+ \omega t }\,,\label{n8+}
\gal
\end{subequations}
where $u_0$, $u_\pm$ are constants.  Since $\nu_+= \nu_-^*$  the second moments and hence the amplitudes are real if $u_+=-u_-^*$. The solution thus reads
\begin{subequations}\label{n12}
\bal
&\aver{x^2}= \frac{1}{2}\left[ u_0 e^{\nu_0 \omega t } -u_+\left(e^{\nu_- \omega t }+e^{\nu_+ \omega t }\right)\right] \,,\label{n9}\\
&\aver{\dot x^2}= \frac{\omega^2}{2}\left[ u_0 e^{\nu_0 \omega t } +u_+\left(e^{\nu_- \omega t }+e^{\nu_+ \omega t }\right)\right]\,,\label{n10}\\
&\aver{x\dot x}= \frac{\omega u_+}{2i}\left( e^{\nu_+ \omega t }-e^{\nu_- \omega t }\right) \,.\label{n11}
\gal
\end{subequations}
When plugged into \eq{n13},\eq{n14} it yields the time-evolution of the average energy and magnetic helicity. I turn now to the analysis of the instabilities. Note, that the instabilities of \eq{n12} are different from those of  \eq{m20}.

%%%%%%
\subsection{The underdamped modes $k> \sigma/2$}\label{sec:n1}

Consider the well conducting chiral matter. Examination of \eq{n12} and \eq{n13} reveals that the only underdamped ($\omega^2>0$)  non-resonant ($c_2>0$) modes that are potentially unstable are those that satisfy $\nu_0\omega>\sigma$, i.e.\ $\omega\alpha^2(c_3-c_2)>2\sigma$. For a  good conductor 
\eq{p3} implies that $c_3-c_2\sim 1/\omega\tau_c$, so that this condition can be rewritten as $k^2\Sigma_\chi^2/\omega^4\tau_c \sigma >1$.  At small $\omega$ one can expand $k$ near $\sigma/2$ to obtain  $\delta k^2<\Sigma_\chi^2/\sigma\tau_c$. This condition, however, contradicts \eq{o5} indicating that no underdamped non-resonant mode of a good conductor is divergent.

At the parametric resonance ($c_2<0$), the instability can arise when either $\nu_0\omega>\sigma$ or 
$\text{Re}\,\nu_\pm \omega>\sigma$. In the former case, using $c_3-c_2\sim \omega \tau_c$ and repeating the arguments of the previous paragraph yields  $\delta k < \Sigma^2_\chi \tau_c$ which contradicts $\omega \gtrsim 1/\tau_c$ as was argued beneath \eq{m21.2}. It actually also contradicts \eq{o5}. In the later case, using $c_3+3c_2\sim -\omega\tau_c$ again implies the contradictory condition $\delta k < \Sigma_\chi^2\tau_c$. 

Thus, the underdamped modes of a well conducting chiral matter are stable. The dissipation beats the growth. 

Turning to the poorly conducting chiral matter we distinguish two cases. (i)  If $\omega \tau_c\gg 1$, then $c_3-c_2\sim 1/\omega \tau_c$ and $\nu_0\omega >\sigma$ implies $\delta k^2 <\Sigma_\chi^2/\sigma\tau_c$ that contradicts \eq{o5}. (ii) If $\omega \tau_c\ll 1$, then according to \eq{p9}  $c_3-c_2\sim \omega \tau_c$ and  $\delta k<\Sigma_\chi^2\tau_c$. 
The consistency with \eq{o5} requires that $\sigma\tau_c<1$ which is allowed for a poor conductor since $\sigma\tau_c\sim e^2/g^4\ll 1$. Thus we conclude that these modes are unstable, provided that the electrical conductivity is sufficiently small $\sigma < 1/\tau_c\sim Tg^4$.   

\begin{figure}[t]
      \includegraphics[height=5cm]{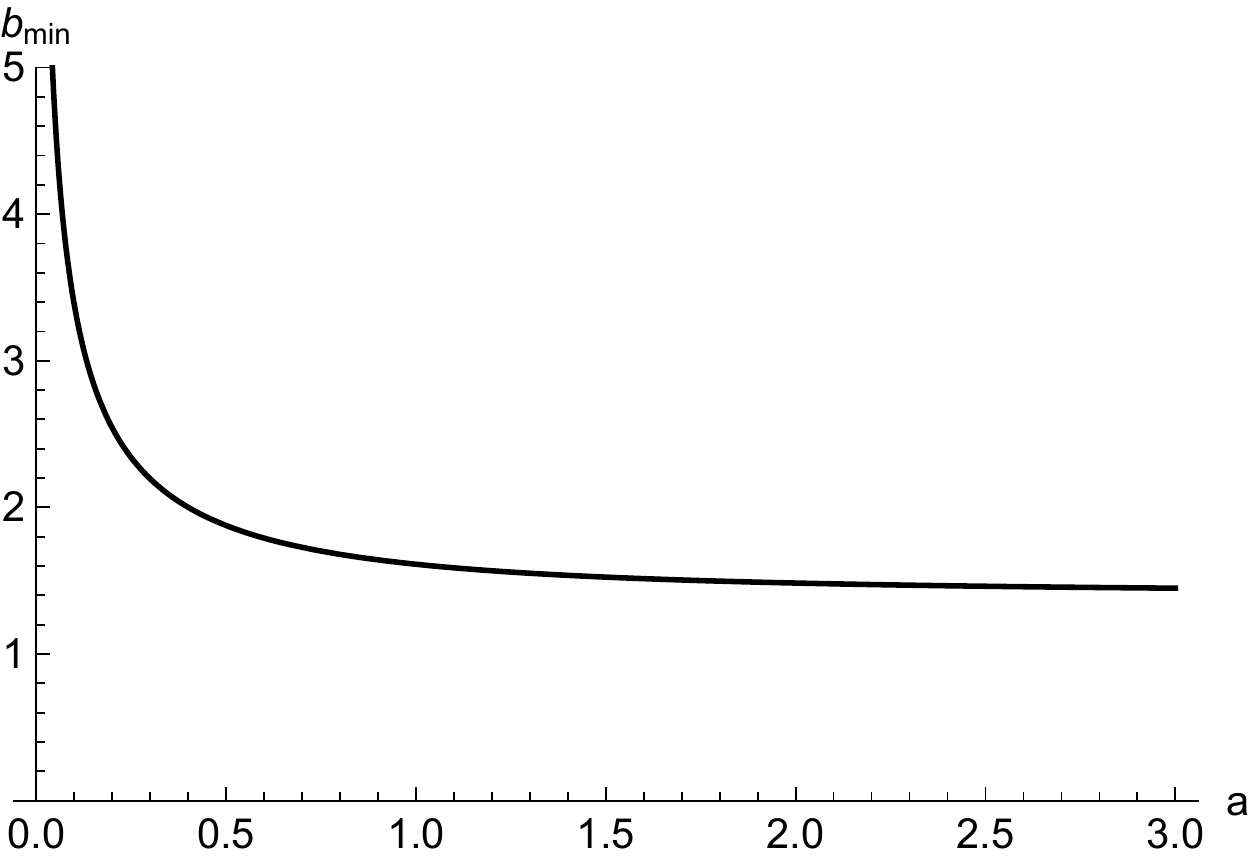} 
  \caption{The minimum values of $b=\Sigma_\chi/\sigma$ at given $a=\sigma\tau_c$ at which the average energy of magnetic field is unstable in a poorly conducting chiral matter.}
\label{fig1}
\end{figure}

The unstable modes of average energy read
\ball{n32}
\aver{\mathcal{E}_{\b k\lambda}}=\frac{k}{2}u_0 e^{\nu_0 k t-\sigma t}= \frac{k}{2}u_0\exp\left\{ \frac{\Sigma_\chi^2}{2k}(c_3-c_2)t-\sigma t\right\}\,.
\gal
To make a more quantitative estimate it is useful to employ the model \eq{p7}. Eq.~\eq{n32} reads 
\ball{n33}
\aver{\mathcal{E}_{\b k\lambda}}=\frac{k}{2}u_0\exp\left\{ \frac{\Sigma_\chi^2}{2k}\frac{2\omega\tau_c}{1+4\omega^2\tau_c^2}t-\sigma t\right\}\,.
\gal
Denote $y=2\omega \tau_c$, $a=\sigma \tau_c$ and $b=\Sigma_\chi/\sigma$. With this notations one can write \eq{n33} as
\ball{n35}
\aver{\mathcal{E}_{\b k\lambda}}=\frac{k}{2}u_0e^{f(y)  \sigma t}\,,\qquad f(y)= \frac{b^2 y}{\sqrt{y^2/a^2+1}(1+y^2)}-1\,.
\gal
The expression in the exponent has a maximum 
\ball{n36}
f(y_0)= \frac{4\sqrt{2}ab^2}{\left[4+a(-a+\sqrt{8+a^2})\right]\sqrt{2+a(a+\sqrt{8+a^2})}}-1
\gal
at $y_0= \frac{1}{2}\sqrt{-a^2+a\sqrt{8+a^2}}$. The instability develops only if $f(y_0)>0$ which gives the minimal value of $b$ as a function of $a$
\ball{n37}
b_\text{min}=a^{-1/2}(1+y_0^2)^{1/2}(1+a^2+2y_0^2)^{1/4}
\gal
shown in \fig{fig1}. $f(y_0)$ monotonically increases with $a$ reaching $b^2/2-1$ at $a\to \infty$. Thus, at large $a$ the condition for instability is $b^2>2$. 

 For example, consider the quark-gluon plasma produced in heavy-ion collisions. It has the electrical conductivity  $\sigma\approx 5$~MeV, and the correlation time $\tau_c\approx 5$~fm so that $a= 0.12$. From $f(y_0)=0$ in \eq{n36} one can find that the corresponding $b_\text{min}= 3.1$. Thus, the instability in the magnetic field energy occurs if the chiral conductivity dispersion is at least of the order of  $\Sigma_\chi= 15$~MeV. 

%%%%%%
%%%%%

Plugging \eq{n12} into \eq{n14} one finds that the average magnetic helicity of each mode has the same instability as the average energy
\ball{n40}
\aver{\mathcal{H}_{\b k\lambda}}=\frac{\lambda u_0}{2}e^{\nu_0 k t-\sigma t}.
\gal
However, since  $\nu_0$ is proportional to $\alpha^2$ and hence helicity independent, the total average magnetic helicity of the magnetic field \eq{n-3} vanishes identically 
\ball{n43}
\aver{\mathcal{H}}=0\,.
\gal
To be sure, this equation does not preclude the magnetic helicity fluctuations. It only indicates that the contributions of the opposite helicity states  to the magnetic helicity  are on average of equal magnitude and have opposite sign.

 %%%%%%
\subsection{The overdamped modes $k< \sigma/2$}
 
In the overdamped case $\omega^2<0$,  a possible instability may comes about if $\text{Re}(\nu_- \omega)>\sigma$. However, expanding at $k\ll \sigma/2$ one finds $\nu_- \omega-\sigma = -\alpha^2 c_1\sigma/4-2k^2/\sigma$ plus imaginary terms. Thus, the overdamped modes are stable.

%%%%%%
\subsection{The critically damped mode $k= \sigma/2$}

Finally, consider the critically damped mode. As explained in \sec{sec:m}  the corresponding equations can be obtained from \eq{n4+} and \eq{n12} be taking $\omega\to \beta$ and $\alpha\to 1$. In this case  $\nu_0|\beta|-\sigma \sim |\beta|^2\tau_c-\sigma\sim (\Sigma_\chi \tau_c-1)\sigma$, where \eq{p9} and \eq{m102} were used. However, $\Sigma_\chi \tau_c\ll 1$. Thus, the critical mode is stable.

%%%%%%%%%
\subsection{Conclusion}

It has been argued in this section that the average field energy becomes unstable and grows exponentially if the chiral matter has small enough electrical conductivity $\sigma < g^4T$ and large dispersion of the chiral conductivity $\Sigma_\chi\gg \sigma$. The more precise expression is given by \eq{n37}. The average magnetic helicity vanishes.

%%%%%%%%%%%%%%%%%%%%%%%%%%
\section{Summary and discussion}\label{sec:s}

This paper addresses the time-evolution of the magnetic field in hot nuclear matter at times much longer than the sphaleron transition time $\tau_c$. In this limit the chiral conductivity $\sigma_\chi$ is treated as a random process. 
By expanding the electromagnetic field in the basis of the circularly polarized waves it is shown that the field amplitudes obey the stochastic harmonic oscillator equation. The relevant observable quantities are statistical averages such as the average magnetic field and average magnetic field energy. 

Using the method of \cite{VanKampen:1975} the equations for the time-evolution of the average amplitude and its second moments were derived and solved.  The solutions indicate that the average magnetic field is exponentially damped, see \eq{m13}. The time-evolution of the average field energy is more complicated and depends on the electrical conductivity of the chiral matter. At high temperatures when the quark-gluon plasma is expected to be highly conductive, the dissipation effects prevent the development of instabilities. However, at temperatures higher but comparable to the critical temperature, the quark-gluon plasma is known to be a poor electric conductor. It is argued that if the electrical conductivity satisfies $\sigma<g^4T$, the average field energy is unstable, provided that the dispersion of the chiral conductivity fulfills the condition $\Sigma_\chi \gg \sigma$. The corresponding unstable modes are $k>\sigma/2$. A more accurate condition is derived in \sec{sec:n1}.

The magnetic helicity of the right and left-handed modes also increases exponentially. However, the total magnetic helicity vanishes identically. This implies that  the conservation of the total helicity cannot tame the chiral instability at later times as it does at early times  \cite{Boyarsky:2011uy,Hirono:2015rla,Kaplan:2016drz,Tuchin:2017vwb}. Combined with the fact that the growth rate of the average field energy is independent of $\lambda$ indicates that this instability is not chiral. 

These results are in a striking contrast with the constant $\sigma_\chi$ approximation which is valid at early times $t\ll \tau_c$ when the sphaleron transitions can be neglected. In that case, the magnetic field itself is unstable and the unstable modes are soft $k<\sigma_\chi$. Moreover, the instability is chiral as it occurs for only one of the helicity modes at a time depending on the sign of $\sigma_\chi$.

%%%%%%%%%%%%%%%%%%%%%%%%%%%%%%%%
\acknowledgments
I wish to  thank Alex Travesset   for an informative discussion. This work  was supported in part by the U.S. Department of Energy under Grant No.\ DE-FG02-87ER40371.

\appendix
\section{Electrodynamics in hot nuclear matter with $CP$-odd domains}\label{appA}

The $CP$-odd domains in the chiral matter can be  described by a pseudo-scalar field $\theta$ whose interactions with the electromagnetic  $F_{\mu\nu}$ and color $G^a_{\mu\nu}$ fields are governed by the Lagrangian \cite{Wilczek:1987mv,Carroll:1989vb,Sikivie:1984yz}
\ball{a1}
\mathcal{L}=\mathcal{L}_\text{QED}+\mathcal{L}_\text{QCD}  -\frac{c_A}{4} \theta   F_{\mu\nu}\tilde F^{\mu\nu}-\frac{c_A'}{4} \theta   G^a_{\mu\nu}\tilde G^{a\mu\nu}  + f^{2}\left[\frac{1}{2}(\partial_\mu \theta)^2-  \frac{1}{2}m^2_\text{ax} \theta^2\right]\,,
\gal
where $\tilde F_{\mu\nu}= \frac{1}{2}\epsilon_{\mu\nu\lambda\rho} F^{\lambda\rho}$ is the dual field tensor,  $c_A$,  $c_A'$ are the QED and QCD anomaly coefficients respectively  and $f$, $m_\text{ax}$ are  constants with mass dimension one. The corresponding equation of motion of the $\theta$-field is 
\ball{a2}
(\partial^2+m_\text{ax}^2)\theta = -\frac{1}{4f^2}\left( c_A' G_{\mu\nu}^a\tilde G^{a\mu\nu} +c_A F_{\mu\nu}\tilde F^{\mu\nu}\right)
\,.
\gal
In the hot nuclear matter the electromagnetic contribution to the topological charge density is presumed to be negligible so that the $\theta$-field dynamics is driven primarily by the topologically non-trivial gluon configurations. The equations of motion  of electromagnetic field  read
\bal
&\partial_\mu F^{\mu\nu}= j^\nu- c_A \tilde F^{\mu\nu}\partial_\mu\theta\,,\label{a5}\\
&\partial_\mu \tilde F^{\mu\nu}= 0\,.\label{a6}
\gal
In the radiation gauge they yield \eq{m1} where  the first derivatives $\partial^\mu\theta$ of the slowly varying field $\theta$ are replaced by their constant domain--average values, in particular  $\sigma_\chi= c_A\dot \theta$ \cite{Tuchin:2019jxd}.

%%%%%%%%%%%%%%%%%%%%%%%%%%%%%%%%%%%%%

\end{document}